\documentclass[10pt,journal,comsoc]{IEEEtran}

\usepackage{cite}
\usepackage{amsmath,amssymb}
\usepackage{graphicx}
\usepackage{booktabs}
\usepackage{algorithm}
\usepackage{algorithmic}
\usepackage{url}

\newif\ifanonymous
\anonymousfalse

\begin{document}

\title{Adaptive Antenna Element Activation for Multiuser Uniform Spherical Arrays}

\ifanonymous
\author{Author~1, Author~2, Author~3, and Author~4}
\else
\author{Ao Li, Cunhua Pan,~\IEEEmembership{Senior Member,~IEEE,}
        Wang Liu, Hong Ren,~\IEEEmembership{Member,~IEEE,}
        and Jiangzhou~Wang,~\IEEEmembership{Fellow,~IEEE}%
\thanks{Ao Li, Cunhua Pan, Wang Liu, Hong Ren, and Jiangzhou Wang are with the National
Mobile Communications Research Laboratory, Southeast University, Nanjing
210096, China(e-mail: aLi@seu.edu.cn; cpan@seu.edu.cn; wangliu@seu.edu.cn; hren@seu.edu.cn; j.z.wang@seu.edu.cn).}%
}
\fi

\maketitle

\begin{abstract}
Uniform spherical arrays provide full-space coverage. However, owing to
directional element patterns, full-array transmission activates many elements
that contribute little to a given user, resulting in increased system overhead
and inefficient power allocation. To address this issue, this letter proposes
an adaptive element activation strategy for multiuser uniform spherical arrays
that reduces the number of physically active elements while satisfying
individual user rate requirements. The proposed strategy constructs a
user-specific candidate sequence according to the angle between each element
boresight and the user direction. At each iteration, it selects, among the
unsatisfied users, the user--element connection that provides the largest
improvement in the aggregate rate deficit. Simulation results show that fewer
than half of the elements are sufficient for a single user to achieve $80\%$
of its full-array rate, while fewer than $70\%$ are sufficient to achieve the
full-array rates of all users in multiuser scenarios. Moreover, the proposed
strategy consistently requires a lower active-element ratio than the baseline
methods across all tested configurations.
\end{abstract}


\section{Introduction}
\IEEEPARstart{E}{xtremely} large-scale multiple-input multiple-output (XL-MIMO)
is regarded as a promising sixth-generation (6G) wireless communication
technology owing to its high array gain and spatial resolution
\cite{Lu2024XLMIMOTutorial}. Conventional uniform linear arrays (ULAs) and
uniform planar arrays (UPAs) are widely used because of their simple structures
and mature implementations. However, off-broadside scanning leads to mainlobe
broadening and reduced angular resolution, while planar arrays may also suffer from
scan-dependent gain loss \cite{Li2023WideAngleReview}. In contrast, uniform
spherical arrays (USAs) offer an attractive array structure owing to their
full-space three-dimensional beam-scanning capability and nearly
direction-independent beamforming gain \cite{Pan2026SphericalReview}.

Full-space coverage by a USA does not imply that every element should serve
every user. Owing to element directivity, elements facing away from a user
typically provide only limited desired-signal gain \cite{Ryoo2021USA}.
Selective element activation can reduce hardware and signal-processing
overhead \cite{Molisch2004Selection}. Therefore, determining suitable active
elements according to user locations is worthy of investigation.

Existing antenna-selection studies have adopted different strategies.
Mendon{\c{c}}a et al. proposed matching-pursuit-based greedy schemes,
including channel-level selection of a prescribed-size common subset for
multiuser transmission \cite{Mendonca2020Greedy}. For spatially
non-stationary XL-MIMO, Ubiali and Abr{\~{a}}o selected the same number of
antennas for each user using an inter-user-coupling-aware channel-power
metric \cite{Ubiali2020VRAntennaSelection}. Their subsequent work used a
cumulative channel-power threshold for flexible antenna and fixed-subarray
selection, allowing user-dependent set sizes
\cite{Ubiali2021FlexibleSelection}. These methods mainly consider ULAs and
do not exploit the distinct element boresights of USAs. For USAs, Ryoo and
Sung confined each beam to a spherical cap and optimized a common
restriction angle for the multiuser sum rate \cite{Ryoo2021USA}. However,
a common angle cannot adapt the activation range to individual users, while
channel-power or sum-rate objectives do not ensure per-user rate
requirements. Adaptive USA element activation under such requirements
therefore remains open.

In this paper, we propose an adaptive element activation scheme for
multiuser USAs that reduces the number of physically active elements while
meeting per-user rate requirements. First, an independent candidate sequence
is constructed for each user according to the angle between each element
boresight and the user direction. At each iteration, only users below their
rate targets propose their next candidates. Shared- and newly activated-element
actions are evaluated by their improvement in the aggregate rate deficit, and
the best candidate is added to the corresponding user's serving set. After
each update, all user rates and the set of users below target are recalculated
until all requirements are met. Thus, each serving-set size adapts to the user
configuration and rate requirements, while the set of physical active elements is the
union of all users' serving sets.

The remainder of this letter is organized as follows. Section~II presents the
system model and problem formulation. Section~III describes the proposed
adaptive element activation strategy. Section~IV presents numerical results
and performance analysis. Section~V concludes the letter.

\section{System Model and Problem Formulation}
\subsection{Uniform Spherical Array and Channel Model}
\begin{figure}[t]
\centering
\includegraphics[width=0.85\columnwidth]{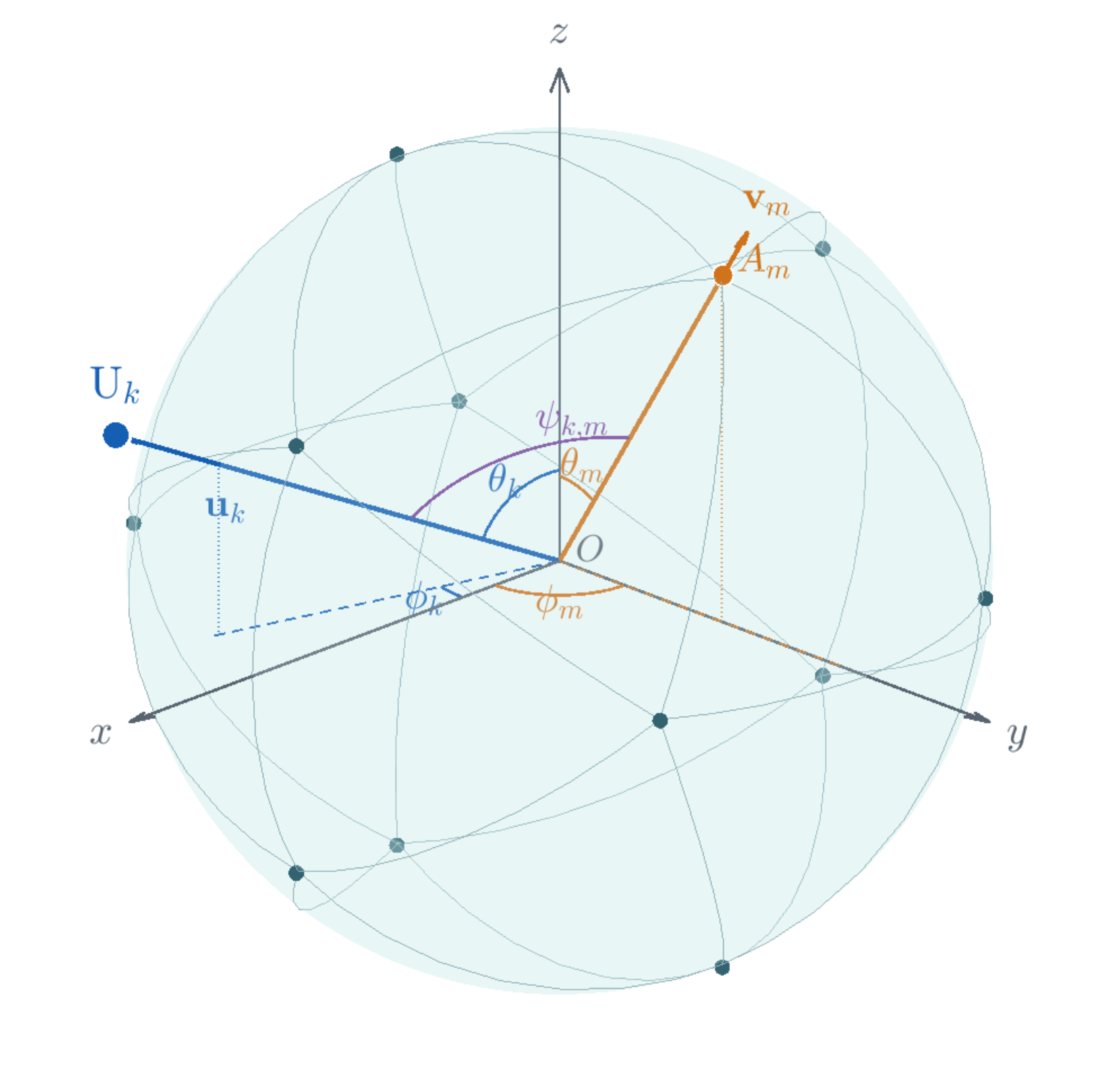}
\caption{Schematic of the spherical array.}
\label{fig:system_geometry}
\end{figure}

As illustrated in Fig.~\ref{fig:system_geometry}, consider a downlink communication
system in which a base station (BS) equipped with an $M$-element USA serves $K$
single-antenna users through $K$ radio-frequency (RF) chains. The BS employs a
fully connected hybrid beamforming architecture that supports selective element
connections. Following \cite{Ryoo2021USA}, the USA is constructed by subdividing
a regular icosahedron and projecting the resulting vertices onto a sphere, with
the element boresights oriented radially outward. The sphere radius is chosen
such that the minimum inter-element distance is $d_{\min}=\lambda/2$.

Taking the array center $O$ as the origin, let $r$ denote the sphere radius.
Element $m$ is denoted by $A_m$, with position vector
$\mathbf p_m=\overrightarrow{OA_m}=r\mathbf v_m$, where $\mathbf v_m$ is the
corresponding unit radial vector. User $k$ is located at $d_k\mathbf u_k$,
where $d_k$ and $\mathbf u_k$ denote its distance and unit direction vector,
respectively. Using the zenith angle $\theta$ and azimuth angle $\phi$, the two
direction vectors are $\mathbf u_k=[\sin\theta_k\cos\phi_k,\allowbreak
\sin\theta_k\sin\phi_k,\allowbreak \cos\theta_k]^{\mathsf T}$ and
$\mathbf v_m=[\sin\theta_m\cos\phi_m,\allowbreak
\sin\theta_m\sin\phi_m,\allowbreak \cos\theta_m]^{\mathsf T}$.

Only line-of-sight (LoS) propagation is considered. With the array center as
the phase reference, the channel coefficient from element $m$ to user $k$ is
expressed as \cite{Ryoo2021USA}
\begin{equation}
h_{k,m}=\frac{\lambda}{4\pi d_k}\sqrt{G(\psi_{k,m})}
\exp\!\left(-j\frac{2\pi}{\lambda}\mathbf p_m^{\mathsf T}\mathbf u_k\right),
\label{eq:channel}
\end{equation}
where $\lambda$ is the carrier wavelength, $G(\psi_{k,m})$ is the element power
gain toward user $k$, and
\begin{equation}
\psi_{k,m}=\arccos\!\left(\mathbf v_m^{\mathsf T}\mathbf u_k\right)
\label{eq:angular_separation}
\end{equation}
is the angular separation between the radial direction of element $m$ and the
direction of user $k$. The normalized element power gain can be expressed as
\begin{equation}
G(\psi)=10^{-\frac{1}{10}\min\left\{
12\left(\frac{\psi}{B_{\mathrm{3dB}}}\right)^2,A_{\max}\right\}},
\label{eq:element_pattern}
\end{equation}
where $B_{\mathrm{3dB}}$ and $A_{\max}$ denote the element half-power beamwidth
and maximum attenuation, respectively.

\subsection{Antenna Element Activation and Rate Constraints}
Let $a_{k,m}\in\{0,1\}$ indicate whether element $m$ serves user $k$. The
serving-element set of user $k$ and the set of physical active elements are,
respectively,
\begin{equation}
\mathcal S_k=\{m:a_{k,m}=1\},\qquad
\mathcal A=\bigcup_{k=1}^{K}\mathcal S_k,
\label{eq:union}
\end{equation}
where a shared element is counted only once in $|\mathcal A|$.

Each active element transmits a total power $P_{\mathrm e}$, equally divided
among the $q_m=\sum_{k=1}^{K}a_{k,m}$ users it serves. For $q_m>0$, its
phase-compensated beamforming weight toward user $k$ is
\begin{equation}
w_{m,k}=a_{k,m}\sqrt{\frac{P_{\mathrm e}}{q_m}}
\exp\!\left(j\frac{2\pi}{\lambda}\mathbf p_m^{\mathsf T}\mathbf u_k\right).
\label{eq:weight}
\end{equation}
If $q_m=0$, element $m$ is switched off. Hence, the total transmit power is
$P_{\mathrm e}|\mathcal A|$.

Let the data symbols be mutually independent, zero-mean, and of unit power,
i.e., $\mathbb E[|s_k|^2]=1$. Define the channel and beamforming vectors as
$\mathbf h_k=[h_{k,1},\ldots,h_{k,M}]^{\mathsf T}$ and
$\mathbf w_k=[w_{1,k},\ldots,w_{M,k}]^{\mathsf T}$, respectively. The received
signal of user $k$ is
\begin{equation}
y_k=\mathbf h_k^{\mathsf T}\mathbf w_k s_k
    +\sum_{j\ne k}\mathbf h_k^{\mathsf T}\mathbf w_j s_j+n_k,
\label{eq:received_signal}
\end{equation}
where $n_k\sim\mathcal{CN}(0,\sigma_k^2)$ is additive Gaussian noise. Treating
the other users' data streams as interference, the achievable rate is
\begin{equation}
R_k=\log_2\!\left(1+
\frac{|\mathbf h_k^{\mathsf T}\mathbf w_k|^2}
{\sum_{j\ne k}|\mathbf h_k^{\mathsf T}\mathbf w_j|^2+\sigma_k^2}\right).
\label{eq:rate}
\end{equation}

\begin{figure*}[!t]
\centering
\includegraphics[width=\textwidth]
  {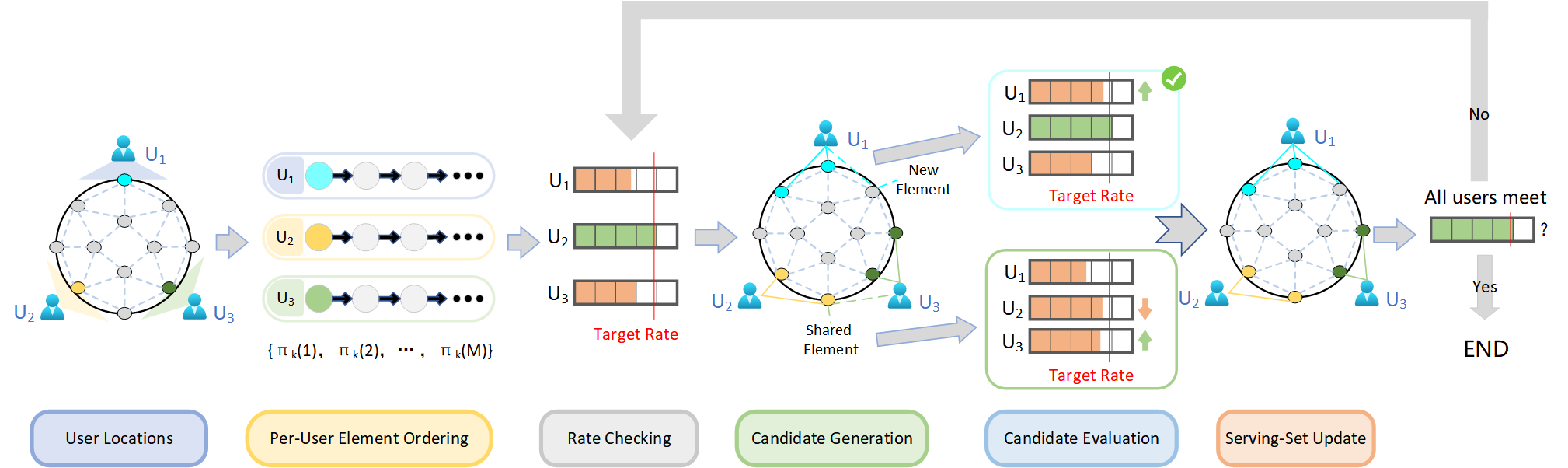}
\caption{Framework of the proposed adaptive antenna element activation strategy.}
\label{fig:activation_framework}
\end{figure*}

Under the full-array baseline, every element serves all users, i.e.,
$a_{k,m}=1$ and $q_m=K$, with beamforming vector
$\mathbf w_k^{\mathrm{full}}$. For a common reference SNR $\gamma_0$, the noise
power of user $k$ is set to
\begin{equation}
\sigma_k^2=\frac{|\mathbf h_k^{\mathsf T}\mathbf w_k^{\mathrm{full}}|^2}
{\gamma_0},
\label{eq:noise_calibration}
\end{equation}
and remains fixed during activation. Substituting
$\mathbf w_k^{\mathrm{full}}$ into \eqref{eq:rate} gives the full-array rate
$R_k^{\mathrm{full}}$.

For a prescribed rate-retention factor $\beta\in(0,1]$, the optimization problem
is
\begin{subequations}\label{eq:problem}
\begin{align}
\underset{\{a_{k,m}\}}{\operatorname{minimize}}\quad
    & |\mathcal A|,\\
\operatorname{subject\ to}\quad
    & R_k\geq \beta R_k^{\mathrm{full}},\quad \forall k,\\
    & a_{k,m}\in\{0,1\},\quad \forall k,m.
\end{align}
\end{subequations}

Since changing a serving set alters power sharing at overlapping elements and
inter-user interference, the user rates are coupled. The next section therefore
develops an adaptive strategy that progressively adds user--element connections.

\section{Proposed Adaptive Activation Strategy}
The overall workflow of the proposed strategy is illustrated in
Fig.~\ref{fig:activation_framework}. It comprises five stages: per-user element ordering,
 rate checking, candidate generation, candidate evaluation, 
 and serving-set update. After the initial ordering, the 
 latter four stages are repeated until all users meet their 
 target rates.

\subsection{Direction-Ordered Candidate Construction}
The radially oriented elements of the USA provide different directional gains
toward a given user. To prioritize the elements that are favorably oriented
toward user $k$, all $M$ elements are sorted in ascending order of
$\psi_{k,m}$. Let
$\boldsymbol{\pi}_k=[\pi_k(1),\pi_k(2),\ldots,\pi_k(M)]$ denote the resulting
user-specific permutation, which satisfies
\begin{equation}
\psi_{k,\pi_k(1)}\leq\psi_{k,\pi_k(2)}\leq\cdots
\leq\psi_{k,\pi_k(M)}.
\label{eq:ordered_sequence}
\end{equation}
The sequence is constructed independently for every user and remains fixed
throughout the activation process.

At iteration $t$, define the set of users that do not yet satisfy their rate
requirements as
\begin{equation}
\mathcal U^{(t)}=\left\{k:R_k^{(t)}<\beta R_k^{\mathrm{full}}\right\}.
\label{eq:unsatisfied_set}
\end{equation}
Only these users are allowed to propose candidate actions. Let
$\ell_k^{(t)}=|\mathcal S_k^{(t)}|$ be the number of elements currently serving
user $k$. The valid candidate-user set and the next candidate element are
\begin{equation}
\mathcal C^{(t)}=\left\{k\in\mathcal U^{(t)}:\ell_k^{(t)}<M\right\},
\qquad
m_k^{(t)}=\pi_k\!\left(\ell_k^{(t)}+1\right).
\label{eq:next_candidate}
\end{equation}
Thus, each unsatisfied user proposes at most one action per iteration. If
$m_k^{(t)}\notin\mathcal A^{(t)}$, the action activates one new physical
element. If $m_k^{(t)}\in\mathcal A^{(t)}$, the element is already active for
another user and the action only adds a new user--element connection. The
latter is referred to as a shared-element action and does not increase
$|\mathcal A^{(t)}|$.

\subsection{Rate-Deficit-Based Action Selection and Iterative Update}
Adding a user--element connection can change both the power assigned to users
sharing that element and the resulting inter-user interference. Consequently,
each candidate must be evaluated according to its effect on all users. Define
the aggregate rate deficit at iteration $t$ as
\begin{equation}
D^{(t)}=\sum_{k=1}^{K}
\left[\beta R_k^{\mathrm{full}}-R_k^{(t)}\right]^+,
\label{eq:deficit}
\end{equation}
where $[x]^+=\max(x,0)$. For the candidate proposed by user $k$, the connection
to $m_k^{(t)}$ is first applied tentatively. The associated power sharing,
beamforming weights, and all user rates are then recalculated, yielding the
trial deficit $D_k^{(t,+)}$. The improvement of this action is
\begin{equation}
\Delta_k^{(t)}=D^{(t)}-D_k^{(t,+)}.
\label{eq:improvement}
\end{equation}

After evaluating all valid candidates, the selected user is
\begin{equation}
k^\star=\underset{k\in\mathcal C^{(t)}}{\arg\max}\;
\Delta_k^{(t)}.
\label{eq:action_selection}
\end{equation}
Thus, shared-element and newly activated-element actions are compared using
the same aggregate-deficit improvement metric. Due to per-element power sharing
and inter-user interference, the aggregate rate deficit may temporarily
increase after adding a user--element connection. If all candidate actions
yield negative improvements, the action with the largest $\Delta_k^{(t)}$ is
still selected because it causes the smallest increase in the aggregate
deficit, thereby allowing the activation process to continue rather than
becoming trapped at the current state.

After selecting $k^\star$, the serving sets are updated as
\begin{equation}
\mathcal S_k^{(t+1)}=\begin{cases}
\mathcal S_k^{(t)}\cup\{m_{k^\star}^{(t)}\}, & k=k^\star,\\
\mathcal S_k^{(t)}, & k\neq k^\star,
\end{cases}
\label{eq:serving_set_update}
\end{equation}
and $\mathcal A^{(t+1)}=\bigcup_{k=1}^{K}\mathcal S_k^{(t+1)}$. The power
allocation and all user rates are subsequently updated, after which
\eqref{eq:unsatisfied_set} is evaluated again. The procedure terminates when
$\mathcal U^{(t)}=\varnothing$.

The complete procedure of the proposed adaptive element activation strategy is
summarized in Algorithm~\ref{alg:activation}.

\subsection{Complexity Discussion}
Each iteration establishes one previously unused user--element connection, and
no accepted connection is removed. Since there are at most $KM$ such
connections, the number of iterations $T$ satisfies $T\leq KM$. In the worst
case, every element serves every user, recovering the full-array configuration;
thus, the rate targets are feasible for $0<\beta\leq1$, and the procedure
terminates after a finite number of iterations.

At each iteration, at most $K$ candidate actions are evaluated. Since one
candidate changes only the $K$ beamforming coefficients associated with one
element, updating the effective-channel matrix and all user rates requires
$\mathcal O(K^2)$ operations. Hence, evaluating at most $K$ candidates per
iteration requires $\mathcal O(K^3)$ operations, and $T$ iterations require
$\mathcal O(K^3T)$ operations. Including the $\mathcal O(KM\log M)$ direction
ordering, the overall complexity is
$\mathcal O(KM\log M+K^3T)$, which is upper bounded by
$\mathcal O(KM\log M+K^4M)$ because $T\leq KM$.

\begin{algorithm}[H]
\caption{Proposed Adaptive USA Element Activation Strategy}
\label{alg:activation}
\begin{algorithmic}[1]
\REQUIRE User-specific ordered sequences $\{\boldsymbol{\pi}_k\}$, target rates
$\{\beta R_k^{\mathrm{full}}\}_{k=1}^{K}$
\ENSURE Serving-element sets $\{\mathcal S_k\}$ and the set of physical active elements
$\mathcal A$
\STATE Initialize $\mathcal S_k\leftarrow\varnothing$ for all $k$ and
$\mathcal A\leftarrow\varnothing$
\STATE Compute all rates and the unsatisfied set $\mathcal U$
\WHILE{$\mathcal U\neq\varnothing$}
  \STATE Form $\mathcal C$ and generate $m_k$ for every $k\in\mathcal C$
  \STATE Tentatively evaluate each action and compute $\Delta_k$
  \STATE $k^\star\leftarrow\arg\max_{k\in\mathcal C}\Delta_k$
  \STATE $\mathcal S_{k^\star}\leftarrow
  \mathcal S_{k^\star}\cup\{m_{k^\star}\}$
  \STATE Update $\mathcal A$, power sharing, all rates, and $\mathcal U$
\ENDWHILE
\end{algorithmic}
\end{algorithm}

\section{Numerical Results}
The far-field LoS channel model described in Section~II is used in the simulations, with
the carrier frequency set to $30$~GHz. Unless otherwise specified, we set
$M=162$, $B_{\mathrm{3dB}}=90^\circ$, $A_{\max}=30$~dB,
$P_{\mathrm e}=0.01$~W, and $\gamma_0=20$~dB. Each result is averaged over 500
random user-location realizations, with the azimuth, elevation, and distance
generated within $[-180^\circ,180^\circ]$, $[-90^\circ,90^\circ]$, and
$[20,50]$~m, respectively.

\begin{figure}[htbp]
\centering
\includegraphics[width=0.95\columnwidth]{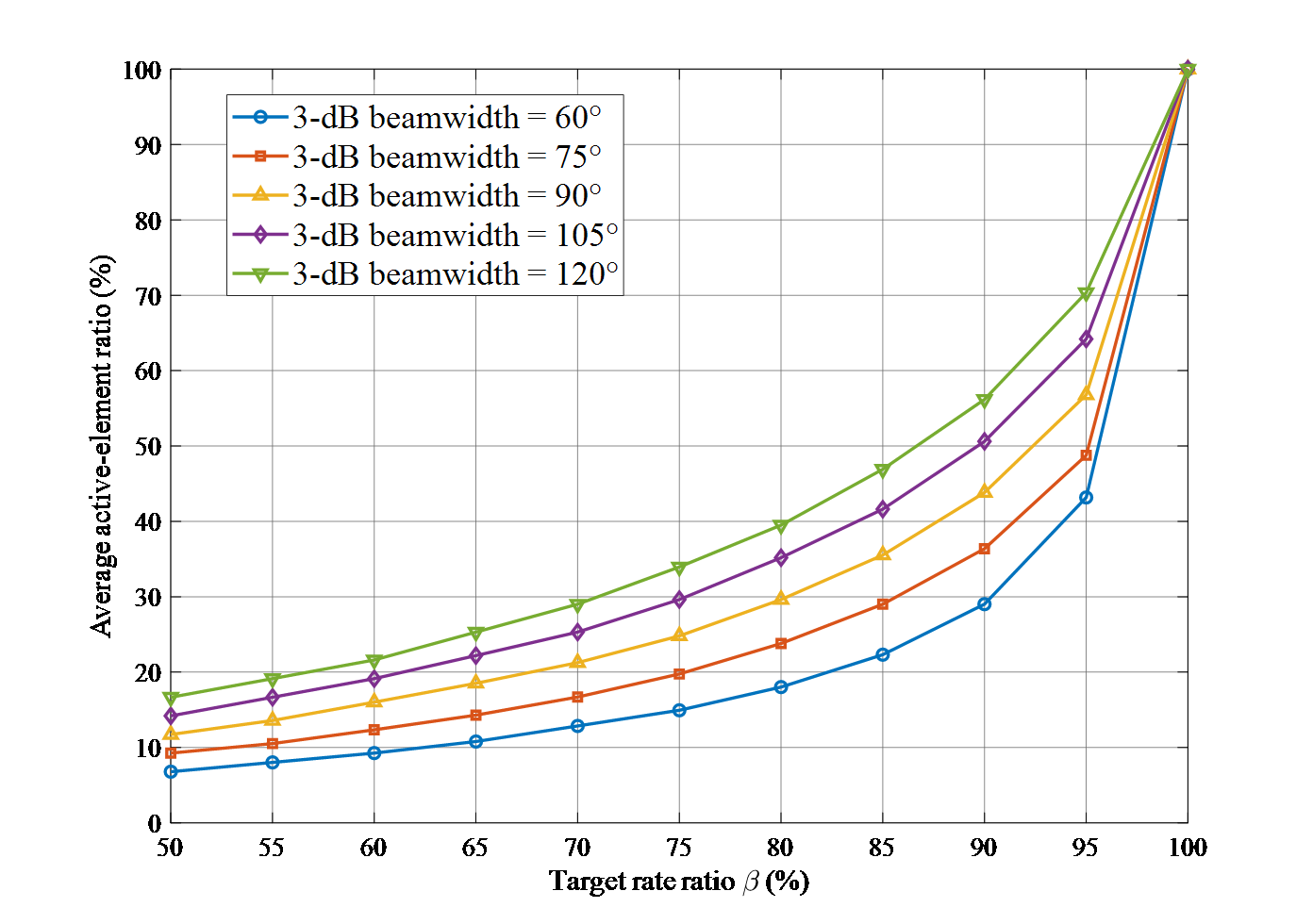}
\caption{Average active-element ratio of the proposed method for different
element beamwidths in the single-user case.}
\label{fig:single_user_activation}
\end{figure}

Fig.~\ref{fig:single_user_activation} compares the required active-element
ratios for different element beamwidths. It can be seen from the figure that the active-element ratio increases
slowly at first and then rapidly as $\beta$ approaches one. In particular,
less than half of the array is sufficient to achieve $80\%$ of the full-array
rate, demonstrating the benefit of selective element activation. For a fixed
$\beta$, the active-element ratio increases with $B_{\mathrm{3dB}}$, since a
wider element beam distributes the useful contribution over more elements and
thus requires a larger active subset.

\begin{figure}[htbp]
\centering
\includegraphics[width=0.95\columnwidth]{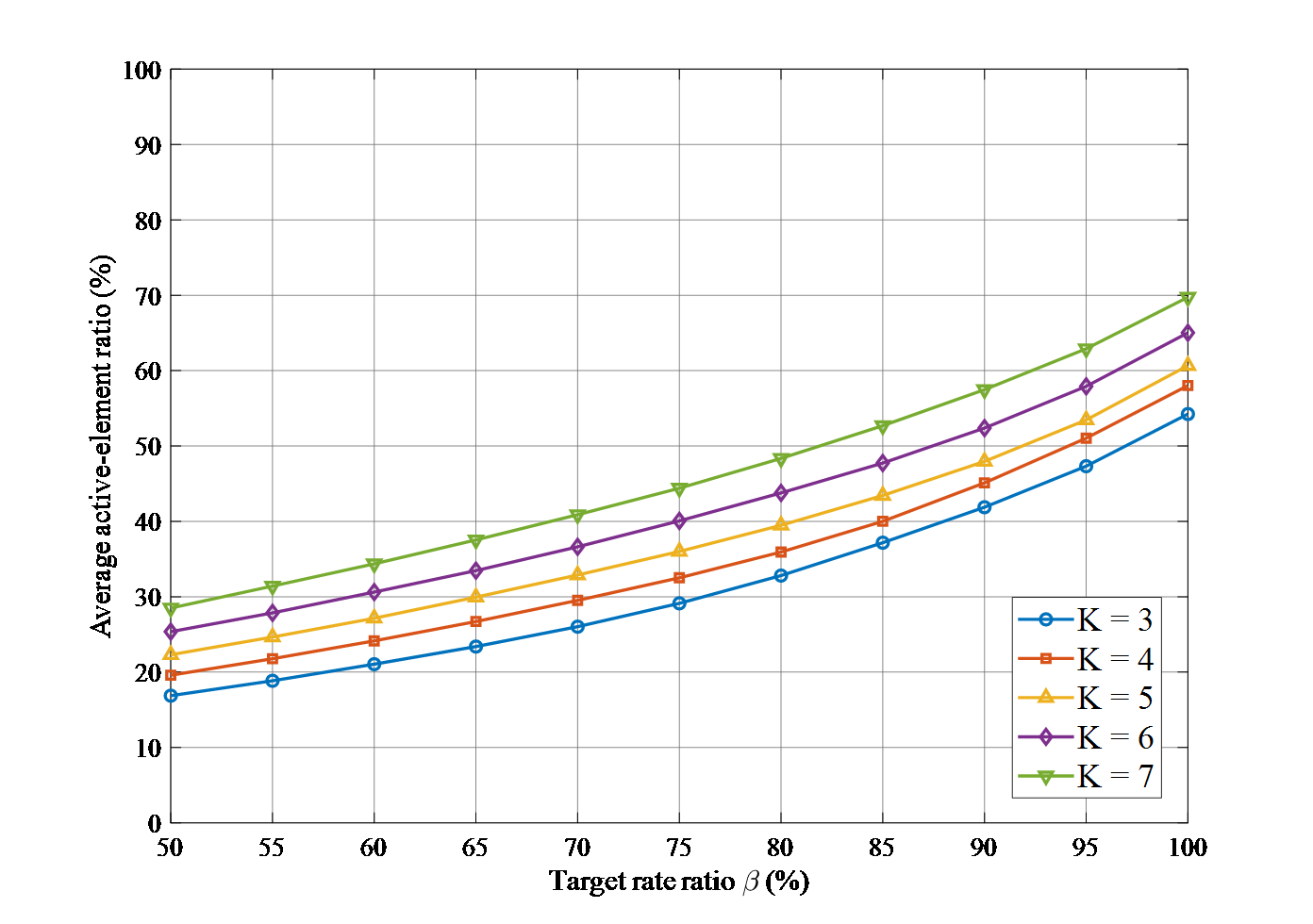}
\caption{Average active-element ratio of the proposed method for different
numbers of users in the multiuser case.}
\label{fig:multiuser_activation}
\end{figure}

\begin{figure*}[htbp]
\centering
\includegraphics[width=0.95\textwidth]{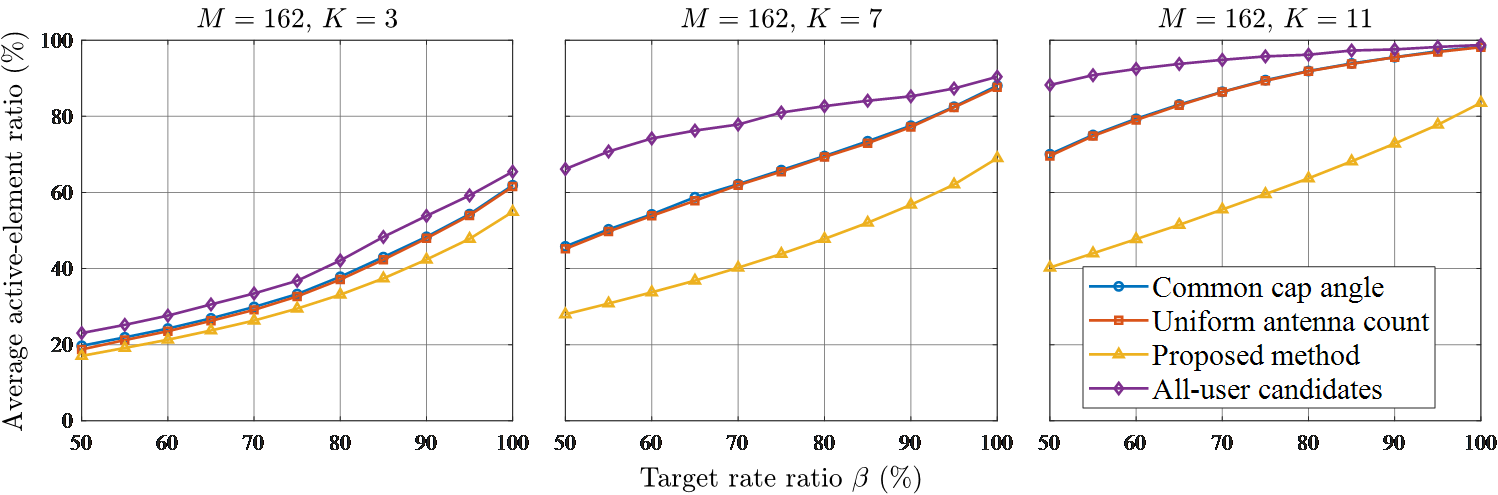}
\caption{Average active-element ratio of different activation methods for
$M=162$ and $K=3$, $7$, and $11$.}
\label{fig:algorithm_comparison}
\end{figure*}

\begin{table*}[htbp]
\centering
\caption{Average runtime at $\beta=0.8$ ($\mathrm{ms}$)}
\label{tab:runtime}
\setlength{\tabcolsep}{4pt}
\begin{tabular}{lccccccccc}
\toprule
& \multicolumn{3}{c}{$M=42$}
& \multicolumn{3}{c}{$M=162$}
& \multicolumn{3}{c}{$M=642$} \\
\cmidrule(lr){2-4}\cmidrule(lr){5-7}\cmidrule(lr){8-10}
Method
& $K=3$ & $K=7$ & $K=11$
& $K=3$ & $K=7$ & $K=11$
& $K=3$ & $K=7$ & $K=11$ \\
\midrule
Common cap angle
& 0.133 & 0.244 & 0.399
& 0.239 & 0.477 & 0.929
& 0.801 & 1.693 & 3.015 \\
Uniform antenna count
& 0.054 & 0.118 & 0.201
& 0.181 & 0.397 & 0.807
& 1.038 & 2.053 & 3.604 \\
Proposed adaptive method
& 0.199 & 0.666 & 1.971
& 0.771 & 1.993 & 5.382
& 4.239 & 11.690 & 24.545 \\
\bottomrule
\end{tabular}
\end{table*}

Fig.~\ref{fig:multiuser_activation} illustrates the effect of the number of
users on the active-element ratio. For a fixed target rate ratio $\beta$, the
required active-element ratio increases with $K$, since more users generally
expand the union of their serving-element sets. Nevertheless, even at
$\beta=1$, only about $54\%$--$68\%$ of the array is activated for $K=3$--$7$,
while all users achieve their full-array rates. This is because, under
full-array transmission, the power of each element is shared among all users,
whereas element selection allows some elements to primarily serve users near
their boresight, thereby reducing inefficient power allocation to users in
mismatched directions.

For fair comparison, the common-cap-angle and uniform-antenna-count methods in
\cite{Ryoo2021USA,Ubiali2020VRAntennaSelection} are adapted to the per-user rate
constraints, together with an ablation scheme:
\begin{itemize}
\setlength{\itemsep}{0pt}
\setlength{\parskip}{0pt}
\setlength{\parsep}{0pt}
\item \emph{Common cap angle:} Each cap is centered on the corresponding user
direction, while all users share the same cap half-angle. Starting from
$1^\circ$, the angle increases in $1^\circ$ steps until all users meet their
rate targets.
\item \emph{Uniform antenna count:} The elements are ordered for each user
according to their directional alignment. The same number is selected for every
user and increased one by one until all rate targets are met.
\item \emph{All-user candidates:} Unlike the proposed method, all users,
including those already satisfying their targets, generate candidates at each
iteration. The action-evaluation and update rules remain unchanged.
\end{itemize}
All methods use the same user locations, channel conditions, and stopping
criterion.

Fig.~\ref{fig:algorithm_comparison} compares the average active-element ratios
of the considered methods for $M=162$ under different $K$ and $\beta$. The
common-cap-angle and uniform-antenna-count benchmark methods yield nearly
overlapping results. Because the elements are approximately uniform over the
sphere, enlarging the common cap is similar to adding the same number of
direction-ordered elements.
In contrast, the proposed method independently adjusts each serving-set size
according to the corresponding user's rate status and therefore consistently
requires a lower active-element ratio. Its advantage over the two benchmark
methods further increases with $K$. With only a few users, the required
serving-set sizes are similar, and a common selection parameter causes limited
waste. As $K$ increases, the user locations, element-sharing relations, and
inter-user interference become more heterogeneous, leading to larger
differences among the required serving-set sizes. The common cap angle or
uniform antenna count is then typically dictated by the most difficult user,
causing overactivation for other users as well as additional power sharing and
interference. By expanding only the serving sets of unsatisfied users and
allowing different set sizes, the proposed method is better suited to complex
multiuser scenarios. Moreover, the all-user-candidate ablation consistently
activates more elements, and its gap from the proposed method widens with $K$.
This confirms that generating candidates for already satisfied users tends to
introduce redundant connections, whereas dynamically updating the unsatisfied
user set more effectively controls physical activation.

Table~\ref{tab:runtime} reports the average runtime at $\beta=0.8$. All methods
become slower as $M$ and $K$ increase because both the per-iteration computation
and the required number of iterations grow. The uniform-antenna-count method is faster
than the common-cap-angle method for smaller arrays but becomes slower at
$M=642$. This is because the common cap angle is scanned in $1^\circ$ steps
with at most 180 trials and is therefore relatively insensitive to $M$, whereas
the uniform antenna count is increased one element at a time and can require up
to $M$ iterations. The proposed method independently selects serving elements
for different users and, in each iteration, evaluates multiple candidate
actions, reallocates element power, and recomputes all user rates. Consequently,
its runtime is the highest and grows more rapidly with $M$ and $K$, increasing
from $0.199$~ms at $(M,K)=(42,3)$ to $24.545$~ms at $(M,K)=(642,11)$. Thus, the
proposed method trades additional computation for a lower active-element ratio.

\section{Conclusion}
This letter studied element activation for multiuser USAs under per-user rate
constraints and proposed an adaptive strategy based on direction ordering and
aggregate rate-deficit improvement. Simulation results showed that fewer than half of
the elements achieve $80\%$ of the full-array rate for a single user, while
about $54\%$--$68\%$ enable all users to attain their full-array rates in
multiuser scenarios. The proposed method consistently outperforms the
baselines across different user configurations, with a more pronounced
advantage at higher user loads, at the cost of increased computation.

\bibliographystyle{IEEEtran}
\bibliography{references}

\end{document}